\documentclass[conference]{IEEEtran}
\IEEEoverridecommandlockouts

\usepackage{cite}
\usepackage{amsmath,amssymb,amsfonts}
\usepackage{algorithmic}
\usepackage[linesnumbered,lined,commentsnumbered]{algorithm2e}
\usepackage{graphicx}
\usepackage{subfigure}
\usepackage{multirow}
\usepackage{textcomp}
\usepackage{xcolor}
\usepackage{adjustbox}

\usepackage[acronym]{glossaries}

\def\BibTeX{{\rm B\kern-.05em{\sc i\kern-.025em b}\kern-.08em
    T\kern-.1667em\lower.7ex\hbox{E}\kern-.125emX}}

\usepackage[normalem]{ulem}

\makeatletter
\newcommand{\linebreakand}{%
  \end{@IEEEauthorhalign}
  \hfill\mbox{}\par
  \mbox{}\hfill\begin{@IEEEauthorhalign}
}
\makeatother

\newacronym{3GPP}{3GPP}{3rd Generation Partnership Project}
\newacronym{5GNR}{5GNR}{{5G} new radio}
\newacronym{5G}{5G}{fifth generation}
\newacronym{5G-PPP}{5G-PPP}{{5G} Infrastructure Public Private Partnership}
\newacronym{5GS}{5GS}{5G system}
\newacronym{AI}{AI}{artificial intelligence}
\newacronym{AGV}{AGV}{automated guided vehicle}
\newacronym{AP}{AP}{access point}
\newacronym{API}{API}{application programming interface}
\newacronym{AR}{AR}{augmented reality}
\newacronym{AT3S}{ATSSS}{access traffic steering, splitting and switching}
\newacronym{BRF}{BRF}{Bayesian recursive filtering}
\newacronym{CDF}{CDF}{cumulative distribution function}
\newacronym{CP}{CP}{control plane}
\newacronym{CPE}{CPE}{customer-premises equipment}
\newacronym{CU}{CU}{centralized unit}
\newacronym{CUPS}{CUPS}{control and user plane separation}
\newacronym{DL}{DL}{downlink}
\newacronym{DU}{DU}{distributed unit}
\newacronym{eAT3S}{eAT3S}{enhanced \gls{AT3S}}
\newacronym{E2E}{E2E}{end-to-end}
\newacronym{GBR}{GBR}{guaranteed bit rate}
\newacronym{gNB}{gNB}{5G base station}
\newacronym{KPI}{KPI}{key performance indicator}
\newacronym{L2}{L2}{layer 2}
\newacronym{LAN}{LAN}{local area network}
\newacronym{LB}{LB}{load balancing}
\newacronym{LED}{LED}{light-emitting diode}
\newacronym{LiFi}{LiFi}{light fidelity}
\newacronym{LOS}{LOS}{line of sight}
\newacronym{MA-PDU}{MA-PDU}{multi-access \gls{PDU}}
\newacronym{MANO}{MANO}{management and network orchestration}
\newacronym{mmWave}{mmWave}{millimeter-wave}
\newacronym{ML}{ML}{machine learning}
\newacronym{MNO}{MNO}{mobile network operator}
\newacronym{MPTCP}{MPTCP}{multi-path TCP}
\newacronym{N3IWF}{N3IWF}{\mbox{non-3GPP} interworking function}
\newacronym{NFV}{NFV}{network function virtualization}
\newacronym{NLOS}{NLOS}{non-\gls{LOS}}
\newacronym{NPN}{NPN}{non-public network}
\newacronym{NWDAF}{NWDAF}{network data analytics function}
\newacronym{O-RAN}{O-RAN}{open {RAN}}
\newacronym{OCC}{OCC}{optical camera communications}
\newacronym{OT}{OT}{operation technology}
\newacronym{PCF}{PCF}{policy control function}
\newacronym{PDU}{PDU}{protocol data unit}
\newacronym{PLC}{PLC}{programmable logic controller}
\newacronym{PLMN}{PLMN}{public land mobile network}
\newacronym{PLMNID}{PLMNID}{\gls{PLMN} identifier}
\newacronym{PLR}{PLR}{packet loss rate}
\newacronym{PMF}{PMF}{performance measurement function}
\newacronym{PNI-NPN}{PNI-NPN}{public network integrated \gls{NPN}}
\newacronym{PRB}{PRB}{physical resource block}
\newacronym{QoS}{QoS}{quality of service}
\newacronym{RAN}{RAN}{Radio Access Network}
\newacronym{RIC}{RIC}{RAN intelligent controller}
\newacronym{RT}{RT}{real-time}
\newacronym{RTT}{RTT}{round trip time}
\newacronym{RSS}{RSS}{received signal strength}
\newacronym{RSSI}{RSSI}{\gls{RSS} indicator}
\newacronym{RSRP}{RSRP}{reference signal received power}
\newacronym{RSRQ}{RSRQ}{reference signal received quality}
\newacronym{SBA}{SBA}{service-based architecture}
\newacronym{SBMA}{SBMA}{service-based management architecture}
\newacronym{SD}{SD}{smallest delay}
\newacronym{SDN}{SDN}{software defined networking}
\newacronym{SINR}{SINR}{signal-to-interference-plus-noise ratio}
\newacronym{SLA}{SLA}{service level agreement}
\newacronym{SNPN}{SNPN}{Stand-alone \gls{NPN}}
\newacronym{SMF}{SMF}{session management function}
\newacronym{SSID}{SSID}{service set identifier}
\newacronym{SSTSD}{SSTSD}{slice service type - slice differentiator}
\newacronym{TNGF}{TNGF}{trusted \mbox{non-3GPP} gateway function}
\newacronym{ToF}{ToF}{time-of-flight}
\newacronym{TS}{TS}{Technical Specification}
\newacronym{TSN}{TSN}{time-sensitive networking}
\newacronym{TDOA}{TDOA}{time difference of arrival}
\newacronym{TWR}{TWR}{two-way ranging}
\newacronym{UDM}{UDM}{unified data management}
\newacronym{UE}{UE}{user equipment}
\newacronym{UL}{UL}{uplink}
\newacronym{UP}{UP}{user plane}
\newacronym{UPF}{UPF}{user plane function}
\newacronym{VLAN}{VLAN}{virtual local area network}
\newacronym{VIM}{VIM}{virtual infrastructure manager}
\newacronym{VR}{VR}{virtual reality}
\newacronym{vRAN}{vRAN}{virtual Radio Access Network}
\newacronym{WAN}{WAN}{wide area network}
\newacronym[longplural={wireless access technologies}]{WAT}{WAT}{wireless access technology}

\begin{document}

\title{Enhanced Access Traffic Steering Splitting Switching with Utility-Based Decisioning
\thanks{This work is supported by the European Commission’s Horizon 2020 research and innovation program under grant agreement No 871428, \mbox{5G-CLARITY} project.}
\thanks{This article was authored by Tezcan Cogalan while he was associated with InterDigital. Tezcan is now working at Samsung Research UK.}
}

\author{\IEEEauthorblockN{Tezcan Cogalan}
\IEEEauthorblockA{\textit{InterDigital}\\
London, UK \\
tezcan.cogalan@interdigital.com}
\and
\IEEEauthorblockN{Morteza Kheirkhah}
\IEEEauthorblockA{\textit{InterDigital}\\
London, UK \\
morteza.kheirkhah@interdigital.com}
\and
\IEEEauthorblockN{Keya Patani}
\IEEEauthorblockA{\textit{InterDigital}\\
London, UK \\
keya.patani@interdigital.com}
\linebreakand
\IEEEauthorblockN{Daniel Camps-Mur}
\IEEEauthorblockA{\textit{i2CAT}\\
Barcelona, Spain \\
daniel.camps@i2cat.net}
\and
\IEEEauthorblockN{Alain Mourad}
\IEEEauthorblockA{\textit{InterDigital}\\
London, UK \\
alain.mourad@interdigital.com}
}

    
\maketitle

\begin{abstract}
Coexistence of 3GPP access networks, namely 4G and 5G, and \mbox{non-3GPP} access networks, namely Wi-Fi, has enabled data offloading from one access network to another to reduce burden on the congested network. Within the evolution of 5G system, the coexistence has been evolved to a multi-access framework where simultaneous transmission on both 3GPP and \mbox{non-3GPP} access networks has been enabled. This framework is named as \gls{AT3S} in 3GPP and is considered as one of the key enablers to provide stringent requirements on data rate and service continuity. In this paper, we introduce an \gls{eAT3S} algorithm where not only instantaneous link performance measurements but also network operator preferences are taken into account to optimize the overall network performance. As part of considered \mbox{non-3GPP} \glspl{WAT}, \gls{LiFi} access networks are also considered alongside with Wi-Fi networks. 
\end{abstract}

\glsresetall

\begin{IEEEkeywords}
5G-Advanced, 5GNR, Wi-Fi, LiFi, multi-connectivity, ATSSS
\end{IEEEkeywords}

\section{Introduction}
Continuous increase on mobile data traffic usage globally has emerged in various reports over the past ten years. In one of the recent reports \cite{ericsson2021}, it is predicted that by 2026, 5G networks will carry more than half of the world's smartphone traffic. While this estimation is given based on global mobile network data traffic, the data traffic generated by Wi-Fi devices is excluded. The same report also notes that the majority of mobile traffic in urban deployments is generated by indoor users. The huge demand for mobile data traffic in the near future can be justified by ongoing digital transformation from small businesses to large ones. One recent obvious example has been seen on Industry 4.0 transformation where it has paved the way for industries to define use cases with stringent requirements for not only data rate but also latency, reliability and service continuity \cite{5g-use-cases}. Another example to note is the demand on immersive experiences where both high data rates and low latency should be achieved mostly on indoor environments.

As the bandwidth, a.k.a resource, is finite and expensive, efficient use of all available resources is the key optimization objective for mobile network operators. Therefore, it is important that different \glspl{WAT} should coexist and complement each other \cite{lte-wifi-twc}. From this perspective, integrating 4G/5G networks, also known as 3GPP networks, with Wi-Fi networks, also named as a \mbox{non-3GPP} network, has been studied in various releases of the 3GPP system. Initially, the integration is used to utilize one \gls{WAT} at a time. With the evolution of the 5G system, the integration of 3GPP and \mbox{non-3GPP} access networks has enabled a multi-access framework where simultaneous transmission on both 3GPP and \mbox{non-3GPP} access networks becomes possible. In 3GPP, this multi-access framework is named as \gls{AT3S} \cite{3gpp-ATSSS-Rel16,3gpp-ATSSS-Rel17,3gpp-ATSSS-Rel18}. As its name states, the framework has enabled switching, splitting and steering of the access traffic onto 3GPP and \mbox{non-3GPP} networks. 

\Gls{AT3S} framework enforces access traffic steering, splitting and switching strategies in the \gls{UPF} and in the \gls{UE}. 
For example, a given traffic steering rule (the steering rule term is used to represent steering, splitting and switching modes) may assign a fraction of ongoing traffic onto the 3GPP network and the rest of the traffic to the \mbox{non-3GPP} network. 
Another example can be an assignment of a priority indicator for one of the access networks to consider it as the first option to steer traffic flows. 

\Gls{AT3S} has been studied in 3GPP in three phases, namely, Release 16, Release 17 and Release 18, as detailed in \mbox{Section \ref{sec:atsss}}. 
In Release 16 \gls{AT3S} \cite{3gpp-ATSSS-Rel16}, a total of four steering modes were defined. 
In Release 17 \gls{AT3S} \cite{3gpp-ATSSS-Rel17}, a new steering mode called \textit{autonomous steering mode} is defined. In this steering mode, the \gls{UE} and the \gls{UPF} can freely and independently decide how to split the traffic across the two accesses.
Release 18 ATSSS \cite{3gpp-ATSSS-Rel18} has been studying objectives like how to support redundant traffic steering which replicates the packets (for both \gls{GBR} and \mbox{non-\gls{GBR}} traffic) on both accesses and how the traffic can be switched between two \mbox{non-3GPP} access paths in the same operator network in a way to enhance the ATSSS feature.

In the Release 16 ATSSS \cite{3gpp-ATSSS-Rel16} workflow, the \gls{UE} and \gls{UPF} cannot flexibly distribute the traffic over 3GPP and \mbox{non-3GPP} accesses according to real-time link status. 
In other words, the traffic distribution is based on a pre-determined weight factor which is provided by the network operator as a \gls{PCF} rule. 
In Release 17 ATSSS \cite{3gpp-ATSSS-Rel17}, as noted, a new autonomous steering mode is proposed as a solution for the noted problem. 
The working principle of this mode is as follows: 
(i) the \gls{PCF}/\gls{SMF} either does not send a pre-determined weight factor or sends an initial weight factor, and 
(ii) the \gls{UE} and the \gls{UPF} dynamically adjust the weight factor for each access on uplink and downlink, respectively. 
Such a procedure has the following drawbacks. 
First, there is no way for the network operators to simultaneously incorporate their preferences that may reflect their charging policies or \glspl{SLA} and to enable the user plane optimization that the autonomous steering mode can achieve. 
Second, as the \gls{UE} and \gls{UPF} use performance measurements on \gls{RTT}, \gls{UL}/\gls{DL} maximum packet loss rate, \gls{UL}/\gls{DL} maximum jitter as well as \gls{UE}’s internal state (e.g., battery level) to adjust the weight factor dynamically, the performance measurements may change from one epoch to another due to increase/decrease of the traffic load on both accesses. Thus, there can be situations where the weight factors of the different accesses change frequently leading to a ping-pong effect i.e., increasing the weight factor in one epoch and decreasing it in the next epoch. 

A solution to overcome the noted first drawback may consider, firstly, the weight factors adjustment by an implementation specific amount \cite{3gpp-ATSSS-Rel17} (stepwise increase/decrease) when one of the access’ link is not within a configured thresholds of a parameter such as signal power. Then, once both accesses are considered as valid (means within the considered thresholds), the weight factors are re-adjusted to the one indicated by the \gls{PCF}. Although, such a solution may prevent too much divergence from the rules indicated by the \gls{PCF}, it does not mean that it guarantees a convergence to the \gls{PCF} rules/weight factor. Also, the proposed stepwise increase/decrease approach either (i) will not provide a full flexibility for the \gls{UE} and the \gls{UPF} as it considers the scenario where only one of the accesses is valid; or (ii) will suffer from the noted ping-pong effect.

As noted in the second drawback, the \gls{UE} and \gls{UPF} decide on the weight factors based on the combination of the current (real-time) link status and the thresholds for several measurements. However, there is no mechanism defined for the \gls{UE} and \gls{UPF} to either improve the overall network performance for a local area of interest that will take into account operator's preferences or \gls{SLA}s.

This paper proposes an \gls{eAT3S} framework that uses a utility-based approach where both the instantaneous link conditions, access network status as well as operator's preference are taken into account. The motivation behind introducing such an enhancement is to mitigate the noted drawbacks of the current \gls{AT3S} in the 3GPP system.
Moreover, in this paper, IEEE 802.11 based \gls{LiFi} network which is an indoor optical wireless communication system \cite{what-is-lifi,hybrid-lifi-wifi,5g-clarity-commag} is considered as part of an integrated \mbox{non-3GPP} access network along with \mbox{Wi-Fi}. Also, the considered underlying system is based on the \mbox{5G-CLARITY} system described in \cite[Fig.1]{5g-clarity-commag}\footnote{Due to page limitation, a detailed description of the system architecture is omitted in this paper. This paper focuses on ATSSS algorithm and its performance evaluation. Hence, interested readers are referred to \cite{5g-clarity-commag} for further details on the system architecture.}.
 
The rest of the paper is organized as follows. \mbox{Section \ref{sec:atsss}} provides details of 3GPP \gls{AT3S} framework in various releases. \mbox{Section \ref{sec:architecture}} introduces considered architecture focusing on the multi-connectivity and our defined \gls{eAT3S} algorithm.
\mbox{Section \ref{sec:eval}} describes the considered simulation scenario and presents the evaluation results of the proposed \gls{eAT3S} framework. Finally, \mbox{Section \ref{sec:conclusions}} concludes the paper.

\section{ATSSS}\label{sec:atsss}
In Release 16 ATSSS \cite{3gpp-ATSSS-Rel16}, a total of four steering modes were defined. 
For \textit{active-standby}, all  traffic of the \gls{MA-PDU} session is sent to one access only, which is called the ``active" access. The other access serves as a ``standby" access and takes traffic only when the active access becomes unavailable. 
For \textit{smallest delay}, the high priority access is the one that can provide the smallest \gls{RTT} performance. \Gls{PMF} can be used to determine the latency of each access link or alternatively \gls{MPTCP} can be used to obtain such latency measurements. 
For \textit{static Load-Balancing}, a weight information element is used to indicate the proportion of the traffic to be forwarded to 3GPP and \mbox{non-3GPP} access networks. This mode is only applicable to \mbox{non-\gls{GBR}} \gls{QoS} flow. 
For \textit{priority-based}, the two accesses are assigned a priority, and all traffic of the \gls{MA-PDU} session is sent to the high priority access. When congestion arises on the high priority access, new data flows (the ``overflow" traffic) are sent to the low priority access. Also, when the high priority access becomes suddenly unavailable, all traffic is forwarded to the other access (low priority). Note should be taken that the definition of a congested link is implementation specific.

Release 17 ATSSS \cite{3gpp-ATSSS-Rel17} introduced a new steering mode called \textit{autonomous steering mode} where the \gls{UE} and the \gls{UPF} can freely and independently decide how to split the traffic across the two accesses when load-balancing steering mode is in use. 
For all steering modes, the network may provide a \gls{UE}-assistance indication, which indicates that 
(a) the \gls{UE} can decide how to distribute the \gls{UL} traffic based on its internal state (e.g. battery level), and 
(b) the \gls{UE} can request from \gls{UPF} to apply the same distribution for the \gls{DL} traffic, and the \gls{UPF} can take the \gls{UE}'s request into account when deciding the \gls{DL} transmission traffic distribution. 
For all steering modes, the \gls{UE} requests from \gls{UPF} to apply the same distribution for the \gls{DL} traffic by using the \gls{PMF} protocol, if available, or another mechanism, if the \gls{PMF} protocol is not available. 
This other mechanism will be determined during the normative phase of the work. For the Load-Balancing steering mode with fixed weights and Priority-based steering mode, it can be possible to apply a threshold condition, which indicates whether a measured parameter is above or below a threshold. 
The measured parameter in a threshold condition may include (a) the \gls{RTT} and (b) the packet loss rate. Also, the threshold conditions will be the same for both 3GPP and \mbox{non-3GPP} accesses since \gls{QoS} requirements are per service.

\begin{figure}
    \centering
    \includegraphics[width=0.85\columnwidth]{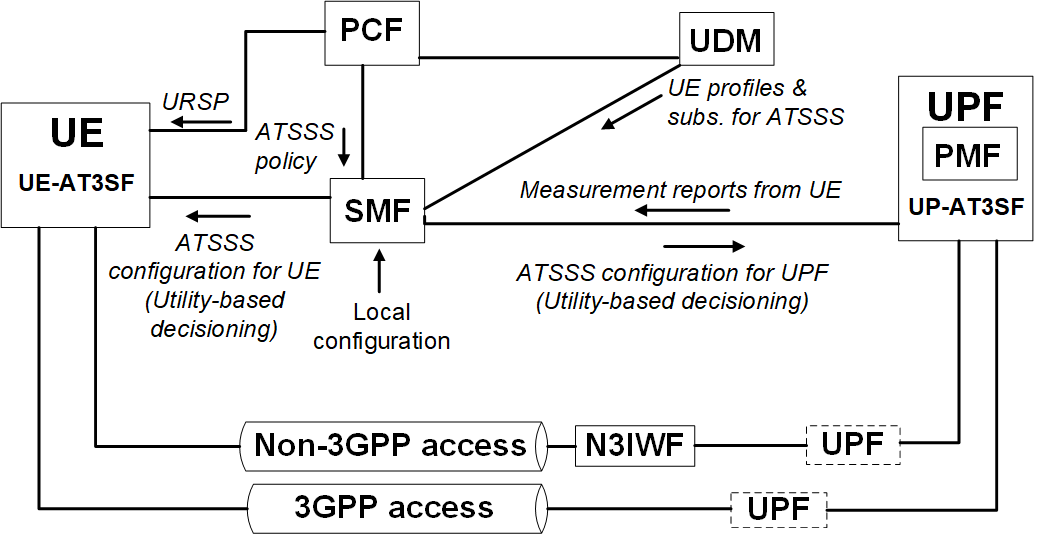}
    \caption{ATSSS execution and interaction between network functions \cite{3gpp-ATSSS-Rel16}. AT3SF stands for ATSSS function; UP stands for user plane; URSP stands for UE route selection policy; N3IWF stands for non-3gpp inter working function.}
    \label{fig:atsss_arch}
\end{figure}

The overall \gls{AT3S} execution procedure is depicted in \mbox{Fig. \ref{fig:atsss_arch}} and can be summarized as follows. First, \gls{PCF} defines an \gls{AT3S} policy based on information from \gls{UDM} such as the \gls{UE} profiles and subscriptions, and then sends \gls{AT3S} policy to \gls{SMF}. Secondly, the \gls{AT3S} policy is configured at the \gls{UE}, more specifically UE-AT3SF where AT3SF stands for \gls{AT3S} function and \gls{SMF}. Then, SMF generates AT3S rules based on the PCF's policy and pushes \gls{AT3S} rules for downlink traffic to \gls{UPF} and \gls{AT3S} rules for uplink traffic to \gls{UE}. Then, \gls{UPF} executes the \gls{AT3S} rule and when downlink data arrives, the \gls{UPF} determines the appropriate access path(s) based on the given \gls{AT3S} rule and sends the downlink data to the UE. \Gls{PMF} performs the path performance measurement for each access path between the \gls{UE} and \gls{UPF}, and reports them to \gls{SMF}. Based on performance measurement results provided by \gls{PMF}, the \gls{SMF} updates the \gls{AT3S} rules and configures updated rules to \gls{UE} and \gls{UPF}.
Although 3GPP does not mandate the specific implementation of the \gls{AT3S} steering function, it allows for different options such as \gls{MPTCP} \cite{mptcp} used in \mbox{5G-CLARITY} project \cite{5g-clarity-commag}. 

The traffic steering strategies enforced via \gls{AT3S} rules are based on predefined values for either all traffic types or some specific traffic type such as UDP or TCP to a specific IP address or port. For example, if the load balancing steering mode is selected, a predefined percentage value has to be written for 3GPP and non-3GPP access networks, such as 20$\%$ for 3GPP and 80$\%$ for non-3GPP. In another example, the priority-based steering mode can be selected to prevent congestion over 3GPP network. Then, high priority is assigned to non-3GPP network to offload the 3GPP network traffic. All these rules are pre-constructed and ordered in a way that as long as a data flow matches a rule, the data flow gets routed according to this rule and the remaining rules are not considered. While traffic is routed according to a specific rule, sudden changes on the network status such as link availability due to channel state fluctuations or link blockage will not be incorporated to traffic routing. 

\section{eAT3S system architecture and algorithm definition} \label{sec:architecture}
The \mbox{5G-CLARITY} system architecture includes virtualized network and application functions (VNFs/VAFs) that can be executed atop the infrastructure including compute, storage and networking fabric. The VNFs implement 3GPP functions, \gls{vRAN}, 5G Core and \gls{UPF}; and provide means to support integration of non-3GPP \glspl{WAT}. The VAFs on the other hand, allow providing domain-specific telemetry services such as \gls{vRAN} telemetry, access network telemetry; and \gls{O-RAN} xApps and service applications. 

The \mbox{5G-CLARITY} architecture considers an integration of \mbox{Wi-Fi} and LiFi access networks to 5G network via \gls{N3IWF}. In \mbox{5G-CLARITY}, although both \mbox{Wi-Fi} and LiFi networks connect to the same \gls{N3IWF}, each network maps to a different IP subnetwork. Thus, it is enabled that separate IP interfaces for \mbox{Wi-Fi}, LiFi and 5G can be managed by \gls{MPTCP} as different paths.

\begin{figure}[ht]
  \centering
  \includegraphics[width=0.8\columnwidth]{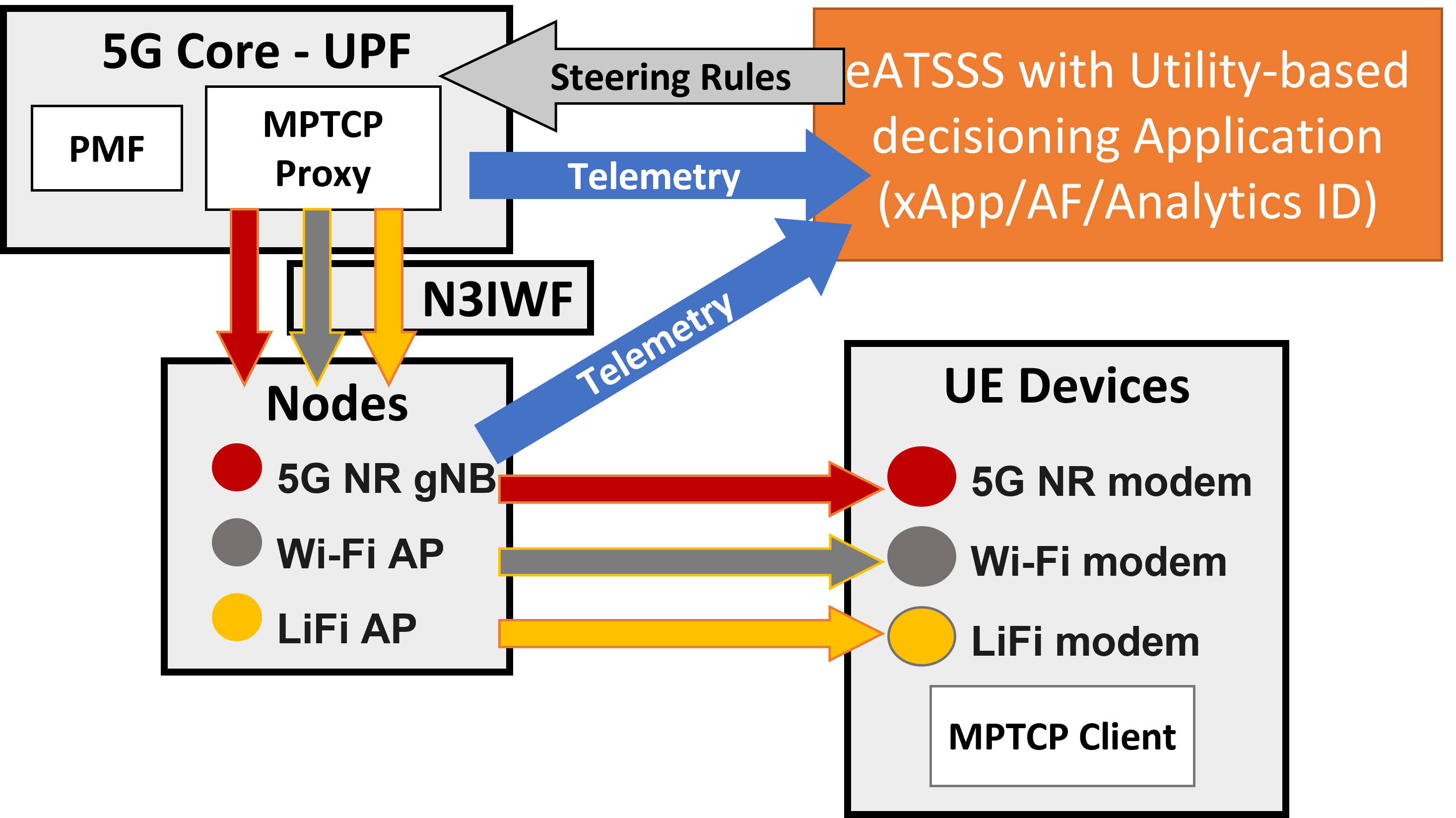}
  \caption{eATSSS execution and interaction between network entities}
  \label{fig:eATSSS_arch}
\end{figure}

In \mbox{Fig. \ref{fig:eATSSS_arch}}, \gls{eAT3S} execution and interaction between network entities are shown. It is worth to note that the flow regarding \gls{PCF}, \gls{SMF} and \gls{UPF} in \mbox{Fig. \ref{fig:atsss_arch}} also exists where \gls{AT3S} rules including operator preference are available at the \gls{eAT3S} with utility-based decisioning application. As it is shown in the figure, the \gls{eAT3S} framework can be implemented as an \gls{O-RAN} xApp, an application function or an analytics as part of 3GPP analytics network function, called \gls{NWDAF}.

\subsection{eATSSS algorithm}
In the proposed utility-based \gls{eAT3S} algorithm, user traffic types are taken into account to both decide on the steering mode selection such as \gls{LB} or \gls{SD} as well as utility function weights. In other words, based on the user requested traffic type, the algorithm decides the steering mode to be used, where different threshold levels are used to identify the satisfaction of various parameters. The threshold levels can be defined by network operator in a static or variable way.
More specifically, if eMBB traffic is requested, the \gls{LB} steering mode will be chosen with a set of specific threshold weights for signal strength, \gls{gNB}/\gls{AP} buffer status, \gls{SINR} and transmission delay, to decide the assigned load level on 3GPP and \mbox{non-3GPP} access networks. In case, URLLC traffic is requested, \gls{SD} mode will be applied with a specific set of threshold weights for the same parameters to decide what access networks out of all available networks should be used to duplicate (or triple if all accesses achieve the utility threshold) the traffic. 


The motivation of using a utility-based algorithm is to unify the steering weight parameters for different service/traffic types and improve the overall network performance. In this paper, a logarithmic utility function in \eqref{eq:log-utility} is used for both eMBB and URLLC traffic types \cite{utility-4g}. In \eqref{eq:log-utility}, $i$ is the telemetry data, namely, \gls{SINR}, buffer or transmission delay; $t$ represents the \gls{WAT}, namely, 5G, \mbox{Wi-Fi} or LiFi; $U_{i}^{t,a}$ is the utility for a given parameter and \gls{WAT} for an active user $a$; $x_{i}^{t,a}$ is the user's measured network-specific telemetry data; and $x_{\mathrm{max}_i}$ is the maximum value considered for a given network-specific telemetry data. 
The utility in \eqref{eq:log-utility} is constructed in a way that $U_{i}^{t,a}(x_{i}^{t,a} \leq 0) = 0$ and $U_{i}^{t,a}(x_{i}^{t,a} \geq x_{\mathrm{max}_i}) = 1$.  
For an eMBB traffic, although the main \gls{KPI} is throughput, transmission delay and buffer status would have impact on the achieved rate when multiple users in the system are competing for the same resources. It is also the same for a URLLC traffic where the main \gls{KPI} is latency and reliability but throughput and buffer status are also important to provide the latency and reliability requirements for a multi-user scenario. Therefore, in the proposed utility-based \gls{eAT3S} algorithm, the weights of \gls{SINR} ($w_{\text{SINR}}$), gNB/AP buffer status ($w_{\text{Buffer}}$) and transmission delay ($w_{\text{Delay}}$) are considered as the weights that can be defined by network operators and can have different values for \gls{LB} and \gls{SD} steering modes. 
Moreover, as each of these parameters would have an impact on the overall performance, an \gls{WAT} that cannot achieve a performance higher than a predefined threshold of $T_{\text{SINR}}$, $T_{\text{Buffer}}$ and $T_{\text{Delay}}$ for \gls{SINR}, buffer status and delay, respectively, is not considered in the utility calculation ($U^{t,a}=0$).

\begin{equation}
    U_{i}^{t,a}(x_i^{t,a}) = \frac{\log \left(1 + w_{i}x_{i}^{t,a}\right)}{\log \left( 1 + w_{i}x_{\mathrm{max}_i} \right)}.
    \label{eq:log-utility}
\end{equation}

Once the utility of each parameter is obtained for each \gls{WAT}, the overall utility is obtained as:
\begin{equation}
    U^{t,a} = U_{\mathrm{SINR}}^{t,a} - U_{\mathrm{Buffer}}^{t,a} - U_{\mathrm{Delay}}^{t,a},
\end{equation}
as higher buffer and delay performance would degrade the overall utility. 
Then, the weight factor assignment for each user for each \gls{WAT}, $W^{t,a}$, is obtained as a percentage on each \gls{WAT} for \gls{LB} in step 12 or a selection of a set of \glspl{WAT} for \gls{SD} steering mode in step 17 in \mbox{Algorithm \ref{alg:eATSSS-algorithm}}.
The execution flow of the proposed \gls{eAT3S} algorithm is provided in \mbox{Algorithm \ref{alg:eATSSS-algorithm}}.

\begin{algorithm}
\caption{Enhanced ATSSS algorithm}\label{alg:eATSSS-algorithm}
\KwData{Telemetry ($x_{i}^{t}$), thresholds ($T_i$), weights ($w_i$)}
\KwResult{Steering weight assignment ($W^{t,a}$)}
$A \gets$ List of active users\;
\While{$A \neq \emptyset$}{
    \For{$t: $ {\upshape 5G, Wi-Fi, LiFi}}{
        \eIf{$x_{\text{SINR,t}}^a > T_{\text{SINR,t}}, ~ x_{\text{Buffer,t}}^a < T_{\text{Buffer,t}}, ~ x_{\text{Delay,t}}^a < T_{\text{Delay,t}}$}{
            {\upshape Obtain} $U_i^{t,a}$ {\upshape by} (1);
        }
        {$U_i^{t,a} = 0;$
        }
    }
    \eIf{{\upshape Steering mode: Load balancing}}{
        \For{$t: $ {\upshape 5G, Wi-Fi, LiFi}}{
        $W^{t,a} =$ $({U_i^{t,a}}/({U^{\mathrm{5G},a}+U^{\mathrm{Wi-Fi},a}+U^{\mathrm{LiFi},a}}))$x$100$;
        }
    }
    {\For{$t: $ {\upshape 5G, Wi-Fi, LiFi}}{
        \If{$U^{t,a} > 0$}{
        $W^{t,a} = 100$;
        }
    }
    }
$A \gets$ $A \setminus a$\;
}
\end{algorithm}

\section{Performance Evaluation}
\label{sec:eval}
\subsection{Scenario description}
\begin{figure}[ht]
  \centering
  \includegraphics[width=0.98\columnwidth]{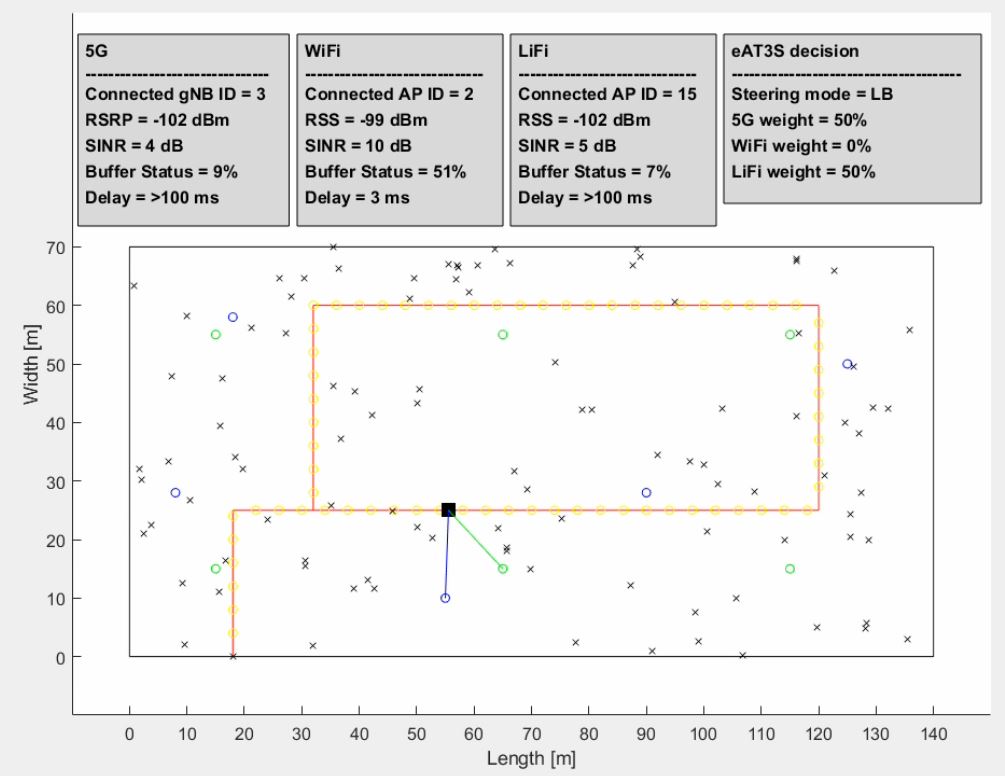}
  \caption{A snapshot from the simulator to depict the simulated factory environment and dynamic change of telemetry data as well as eAT3S steering decisions. \gls{UE}/CPE $x$--$y$ axis coordinates are uniformly distributed.}
  \label{fig:snapshotLB}
\end{figure}

In order to evaluate the performance of the proposed utility-based \gls{eAT3S} algorithm, an \mbox{Industry 4.0} scenario based on a real factory layout, depicted in \mbox{Fig. \ref{fig:snapshotLB}} is chosen. The considered scenario is aligned with a use case defined in the \mbox{5G-CLARITY} project \cite{5gclarity-D5.1}.
The scenario includes the deployment of three different \glspl{WAT}; \mbox{Wi-Fi} at \mbox{5 GHz}, \gls{LiFi} at \mbox{337 THz} and \gls{5GNR} at \mbox{3.5 GHz}. The \mbox{Wi-Fi}, LiFi and \gls{5GNR} systems use the channel bandwidth values of $80$, $20$ and $80$ MHz, respectively. 
In the simulation scenario, an \gls{AGV} that follows the path shown in red in the figure and 100 other users randomly distributed in the considered factory environment exist. 
The randomly distributed 100 users are considered as static users that include factory workers, connected production machinery and sensors. Along the AGV’s path, each \gls{WAT} has a different level of coverage and provides different signal levels from various access nodes. 
In the simulation environment shown in \mbox{Fig. \ref{fig:snapshotLB}}, there are six 5G NR gNBs (shown by green circles) deployed based on factory scenario defined in TR 38.901 \cite{3gpp-tr-38901}, five Wi-Fi APs (shown by blue circles), and LiFi APs (shown by yellow circles along the AGV’s path) deployed with a inter-site distance of 3 meters along the AGV’s predefined path.
In simulations, all users are equipped with a \gls{UE} or \gls{CPE} that provides network connectivity to the all three \glspl{WAT}. 
With the aid of mobility of the \gls{AGV} (with a speed of $0.7$ {m/s}) and $100$ other users that are modeled with a Poisson probabilistic traffic arrivals defined in \mbox{Section \ref{sec:poisson}} as well as a random spatial traffic distribution based on a Zipf popularity distribution defined in \mbox{Section \ref{sec:zipf}}, the proposed \gls{eAT3S} algorithm is evaluated in a realistic environment.

\subsection{Poisson Traffic Model}\label{sec:poisson}
The \gls{UE} requests are modeled as a Poisson arrival process where the sequence of inter-arrival times for \gls{UE} requests are independent and identically distributed (i.i.d) random variables. The Poisson distribution probability mass function $P_R$ of a $r$ number of requests/events in a given time interval $T$ is given as:
\begin{equation}
P_R = \left( \left( {\lambda T} \right) ^{r} e^{-\lambda T} \right) / {r!}    
\end{equation}
where ${\lambda}$ is the Poisson distribution parameter, which is also known as the average number of requests/events during a unit of time; and $(.)!$ is the factorial operation. In the considered system model, a randomly selected \gls{UE} is activated in each Poisson-based request arrival event, and a content $f$ is requested based on the considered Zipf content popularity distribution $P_F$. 

\subsection{Requested content details}\label{sec:zipf}
To enable a random spatial traffic distribution among the deployed APs/gNBs to mimic more realistic network deployment, different content sizes, different popularities among available contents and Poisson arrival process for \gls{UE} requests are considered.

Regarding the content popularity, the popularity within the library of $N_F$ files is characterized as a Zipf distribution \cite{zipf-main,zipf-youtube}. Therefore, the probability $P_F$ of a file $f \in \{1,\ldots, N_F\}$ being requested can be written as:
\begin{equation}
    P_F (f) = Zf^{-\alpha_z},
\end{equation}
where ${\alpha_z}$ is the Zipf parameter that characterizes the probability of content reuse; and $Z=1 ⁄ \sum_{i=1}^{N_F}i^{-\alpha_z}$ is a constant for a given number of files in the library and ${\alpha_z}$. 
When ${\alpha_z}$ is chosen as a value larger than zero, the most popular contents have a lot higher chance to be requested. The content popularity distribution is only used to decide what content should be considered for a given user request

\begin{figure}[ht]
  \centering
  \includegraphics[width=0.8\columnwidth]{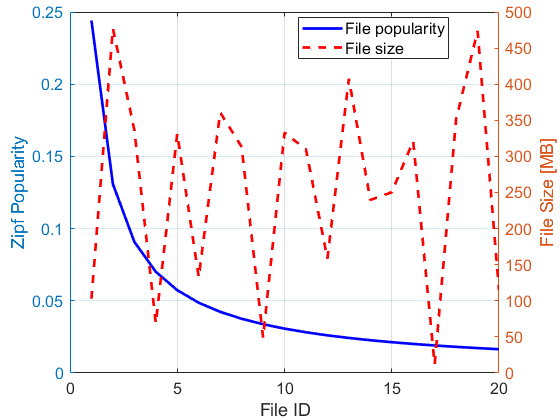}
  \caption{File popularity distribution and file size.}
  \label{fig:contentDetails}
\end{figure}

In addition to content popularity distribution, a random distribution is considered for the file size. This is independent from the content popularity. Each file $f$ in the library of $N_F$ files has a randomly allocated file size as shown in \mbox{Fig. \ref{fig:contentDetails}}.

\subsection{Multi-connectivity evaluation}
\label{subsec:eval_multicon}
\mbox{Fig. \ref{fig:snapshotLB}} shows an instance from a custom system level simulator that mimics the factory environment defined in D5.1 \cite{5gclarity-D5.1}. 
While the AGV moves along the path, the system level simulator provides telemetry readings such as AGV’s connected access node, its received signal power, \gls{SINR}, access node’s buffer status for the downlink traffic and \gls{UE}-specific transmission delay measurements from all three \glspl{WAT}. 
The proposed \gls{eAT3S} algorithm makes use of the available telemetry data. 
The used scheduling type and allocated weight information is also captured in \mbox{Fig. \ref{fig:snapshotLB}}. 
For the \gls{LB} steering mode, the steering score weights of \gls{SINR} ($w_{\text{SINR}}$), gNB/AP buffer status ($w_{\text{Buffer}}$) and transmission delay ($w_{\text{Delay}}$) are set to 1, 0.7 and 0.2, respectively. 
For the \gls{SD} steering mode, $w_{\text{SINR}}$, $w_{\text{Buffer}}$ and $w_{\text{Delay}}$ are set to 0.5, 0.2 and 1, respectively. 
Therefore, it can be expected that for the same telemetry readings for the considered parameters, the utility assessment would be different. 
For example, for the given instance in the figure, the utility metric of the \gls{LB} is different from the \gls{SD}. 
In the considered instance, Wi-Fi access is not considered for the \gls{LB} due to mainly its buffer status. 
However, it may be considered for the \gls{SD} as it has the minimum delay performance among the available \glspl{WAT}. 
Hence, weight allocation and \gls{WAT} selection can differ for each steering mode, depending on the operator’s preference on the steering scoring weights as well as threshold value to consider any of the \gls{WAT} as a candidate for steering decisions. 

\begin{figure}[ht]
  \centering
  \includegraphics[width=0.98\columnwidth]{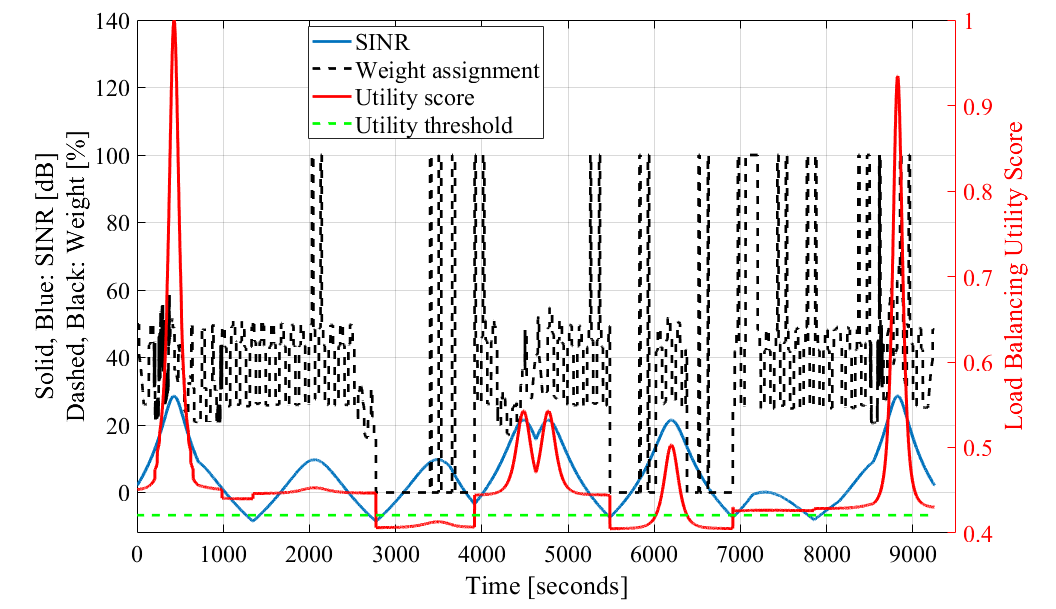}
  \caption{Comparison of 5G \gls{SINR}, weight assignment and steering score for load balancing steering mode when a steering threshold value of 0.6 is used.}
  \label{fig:SINR-LB}
\end{figure}

\mbox{Fig. \ref{fig:SINR-LB}} shows a comparison of 5G \gls{SINR}, weight assignment and utility score for the \gls{LB} steering mode for the whole journey of the AGV when a normalized threshold value of $0.6$ is used for all measurement parameters. The AGV’s traffic is considered as eMBB. The figure shows that the utility scoring of the \gls{LB} steering mode mostly follows the \gls{SINR} performance as shown at $t=300s \rightarrow 700s$, $t=4300s \rightarrow 5000s$, $t=6000s \rightarrow 6500s$ and $t=8600s \rightarrow 9000s$. This is due to $w_{\text{SINR}} = 1$ in the utility scoring. The figure also shows that on some occasions, having a good \gls{SINR} performance is not enough to steer traffic onto the \gls{WAT}. For example, for a period from 3000s to 3500s, although the \gls{SINR} changes from 0 dB to 10 dB, the utility score is not high enough. Hence, the \gls{LB} steering mode does not steer any traffic onto 5G during that period. 

\mbox{Table \ref{tab:compare-LB-SD}} compares the decisioning of both steering modes for the same AGV position. It is worth to note that, although the radio conditions/telemetry is the same for different simulation runs with different traffic type requested by the AGV (eMBB for \gls{LB} and URLLC for \gls{SD}), the buffer status and transmission delay performance are different. This is due to the Poisson-based user request arrival as well as popularity-based content selection. In each run, the user requests and selected content are generated based on the described procedures. In the simulation time of 00:59, the \gls{LB} steering mode divides the access traffic in two and assigns it to 5G and LiFi. However, at the same simulation time, at the same AGV position, the \gls{SD} steering mode only selects \mbox{Wi-Fi} to transmit URLLC traffic as it achieves a transmission delay of 10ms whereas 5G and LiFi achieve 89ms and \textgreater100ms, respectively. Moreover, although \mbox{Wi-Fi} achieves a higher \gls{SINR} performance compared to 5G and LiFi for the same AGV position, its buffer status degrades the utility score of \mbox{Wi-Fi} and the \gls{LB} steering mode discards \mbox{Wi-Fi} in the weight assignment decisioning. Similar conclusions can be derived for the other considered simulation times/AGV positions.

\begin{table}[ht]
\caption{Comparison of decisioning of LB and SD steering modes for the same AGV Position}
\resizebox{\columnwidth}{!}{%
\begin{tabular}{|c|c|ccc|ccc|}
\hline
\multirow{2}{*}{\textbf{Instance}} & \multirow{2}{*}{\textbf{Telemetry}} & \multicolumn{3}{c|}{\textbf{LB}}                                                                 & \multicolumn{3}{c|}{\textbf{SD}}                                                    \\ \cline{3-8} 
                          &                            & \multicolumn{1}{c|}{\textbf{5G}}                & \multicolumn{1}{c|}{\textbf{Wi-Fi}} & \textbf{LiFi}              & \multicolumn{1}{c|}{\textbf{5G}}   & \multicolumn{1}{c|}{\textbf{Wi-Fi}} & \textbf{LiFi}              \\ \hline
\multirow{6}{*}{1}    & Cell ID                    & \multicolumn{1}{c|}{1}                 & \multicolumn{1}{c|}{1}     & 4                 & \multicolumn{1}{c|}{1}    & \multicolumn{1}{c|}{1}     & 4                 \\ \cline{2-8} 
                          & RSRP/RSS {[}dBm{]}         & \multicolumn{1}{c|}{-77}               & \multicolumn{1}{c|}{-101}  & -100              & \multicolumn{1}{c|}{-77}  & \multicolumn{1}{c|}{-101}  & -100              \\ \cline{2-8} 
                          & SINR {[}dB{]}              & \multicolumn{1}{c|}{29}                & \multicolumn{1}{c|}{9}     & 9                 & \multicolumn{1}{c|}{29}   & \multicolumn{1}{c|}{9}     & 9                 \\ \cline{2-8} 
                          & Buffer Status {[}\%{]}     & \multicolumn{1}{c|}{8}                 & \multicolumn{1}{c|}{27}    & 8                 & \multicolumn{1}{c|}{17}   & \multicolumn{1}{c|}{64}    & 8                 \\ \cline{2-8} 
                          & Delay {[}ms{]}             & \multicolumn{1}{c|}{14}                & \multicolumn{1}{c|}{6}     & 2                 & \multicolumn{1}{c|}{18}   & \multicolumn{1}{c|}{9}     & 2                 \\ \cline{2-8} 
                          & Weight {[}\%{]}/Selection  & \multicolumn{1}{c|}{49}                & \multicolumn{1}{c|}{22}    & 29                & \multicolumn{1}{c|}{1}    & \multicolumn{1}{c|}{1}     & 1                 \\ \hline
\multirow{6}{*}{2}    & Cell ID                    & \multicolumn{1}{c|}{3}                 & \multicolumn{1}{c|}{2}     & 15                & \multicolumn{1}{c|}{3}    & \multicolumn{1}{c|}{2}     & 15                \\ \cline{2-8} 
                          & RSRP/RSS {[}dBm{]}         & \multicolumn{1}{c|}{-102}              & \multicolumn{1}{c|}{-99}   & -102              & \multicolumn{1}{c|}{-102} & \multicolumn{1}{c|}{-99}   & -102              \\ \cline{2-8} 
                          & \gls{SINR} {[}dB{]}              & \multicolumn{1}{c|}{4}                 & \multicolumn{1}{c|}{10}    & 5                 & \multicolumn{1}{c|}{4}    & \multicolumn{1}{c|}{10}    & 5                 \\ \cline{2-8} 
                          & Buffer Status {[}\%{]}     & \multicolumn{1}{c|}{9}                 & \multicolumn{1}{c|}{51}    & 7                 & \multicolumn{1}{c|}{14}   & \multicolumn{1}{c|}{65}    & 8                 \\ \cline{2-8} 
                          & Delay {[}ms{]}             & \multicolumn{1}{c|}{\textgreater{}100} & \multicolumn{1}{c|}{3}     & \textgreater{}100 & \multicolumn{1}{c|}{89}   & \multicolumn{1}{c|}{10}    & \textgreater{}100 \\ \cline{2-8} 
                          & Weight {[}\%{]}/Selection  & \multicolumn{1}{c|}{50}                & \multicolumn{1}{c|}{0}     & 50                & \multicolumn{1}{c|}{0}    & \multicolumn{1}{c|}{1}     & 0                 \\ \hline
\multirow{6}{*}{3}    & Cell ID                    & \multicolumn{1}{c|}{6}                 & \multicolumn{1}{c|}{4}     & 37                & \multicolumn{1}{c|}{6}    & \multicolumn{1}{c|}{4}     & 37                \\ \cline{2-8} 
                          & RSRP/RSS {[}dBm{]}         & \multicolumn{1}{c|}{-90}               & \multicolumn{1}{c|}{-79}   & -100              & \multicolumn{1}{c|}{-90}  & \multicolumn{1}{c|}{-79}   & -100              \\ \cline{2-8} 
                          & \gls{SINR} {[}dB{]}              & \multicolumn{1}{c|}{16}                & \multicolumn{1}{c|}{30}    & 10                & \multicolumn{1}{c|}{16}   & \multicolumn{1}{c|}{30}    & 10                \\ \cline{2-8} 
                          & Buffer Status {[}\%{]}     & \multicolumn{1}{c|}{10}                & \multicolumn{1}{c|}{68}    & 10                & \multicolumn{1}{c|}{21}   & \multicolumn{1}{c|}{100}   & 7                 \\ \cline{2-8} 
                          & Delay {[}ms{]}             & \multicolumn{1}{c|}{41}                & \multicolumn{1}{c|}{3}     & 3                 & \multicolumn{1}{c|}{26}   & \multicolumn{1}{c|}{4}     & 2                 \\ \cline{2-8} 
                          & Weight {[}\%{]}/Selection  & \multicolumn{1}{c|}{23}                & \multicolumn{1}{c|}{48}    & 28                & \multicolumn{1}{c|}{0}    & \multicolumn{1}{c|}{1}     & 1                 \\ \hline
\end{tabular}
}
\label{tab:compare-LB-SD}
\end{table}

\section{Conclusions}
\label{sec:conclusions}
It is shown that our proposed \gls{eAT3S} algorithm enables the inclusion of multiple network parameters in the steering decisioning. It enables private/public network operators to derive user or environment specific network policy configurations to efficiently utilize 3GPP and \mbox{non-3GPP} networks. The structure of the proposed utility-based \gls{eAT3S} algorithm also enables an inclusion of machine learning-based approaches that may predict the performance of different network parameters and proactively update the steering weights/\gls{WAT} selections to further improve the user \gls{QoS}/QoE. We leave the definition of machine learning-based \gls{eAT3S} policies as future work.

\end{document}